# Photoresponse of Natural van der Waals Heterostructures

Kyle Ray[¶], Alexander E. Yore[¶], Tong Mou[†], Sauraj Jha[¶], K.K.H. Smithe[‡], Bin Wang[†], Eric Pop[‡] and A.K.M. Newaz[¶]

[¶]Department of Physics and Astronomy, San Francisco State University, San Francisco, California 94132, United States,

[†]School of Chemical, Biological and Materials Engineering, University of Oklahoma, Norman, Oklahoma 73019,

[‡]Department of Electrical Engineering, Stanford University, Stanford, California 94305.

(Date May 9, 2017)

Abstract

Van der Waals (vdW) heterostructures consisting of two dimensional materials offer a platform to obtain material by design and are very attractive owing to novel electronic states. Research on 2D van der Waals heterostructures (vdWH) has so far been focused on fabricating individually stacked atomically thin unary or binary crystals. Such systems include graphene (Gr), hexagonal boron nitride (*h*-BN) and member of the transition metal dichalcogenides family. Here we present our experimental study of the opto-electronic properties of a naturally occurring vdWH, known as Franckeite, which is a complex layered crystal composed of lead, tin, antimony, iron and sulfur. We present here that thin film franckeite (60 nm < *d* < 100 nm) behave as narrow band gap semiconductor demonstrating a wide band photoresponse. We have observed the band-edge transition at ~ 1500 nm (~830 meV) and high external quantum efficiency (*EQE*~3%) at room temperature. Laser power resolved and temperature resolved photocurrent measurements reveal that the photo-carrier generation and recombination are dominated by continuously distributed trap states within the band gap. To understand wavelength resolved photocurrent, we also calculated the optical absorption properties via density functional theory. Finally, we have shown that the device has fast photoresponse with rise time as fast as ~ 1 ms. Our study provides a fundamental understanding of the optoelectronic behavior in a complex naturally occurring vdWH and can open up the possibilities of producing new type of nanoscale optoelectronic devices with tailored properties.

Van der Waals (vdW) heterostructures consisting of two dimensional (2D) materials such as graphene, transition metal dichalcogenides (TMDs), and hexagonal boron nitride (hBN) have gained tremendous attention recently owing to novel electronic states and applications that may be realized in these vertical heterostructures.[1-3] Research on 2D van der Waals heterostructures (vdWH) has so far been focused on layer by layer stacking of these layered materials in the lab, forming synthetic heterostructure materials.[1, 4] This artificial fabrication of vdWH also has major limitations; first, there is little control over crystal orientation while stacking on top of each other and second, the sample can suffer damage by unwanted air bubbles or adsorbates between the stacked layers.

Here we present our experimental study of the photoresponse behavior of thermodynamically stable naturally occurring vdWH, mixed-metal sulfide franckeite, a complex layered crystal composed of lead, tin, antimony, iron and sulfur. Franckeite is formed by alternating stacking of pseudohexagonal (H) layer, e.g. tin disulfide ($SnS_2$), and pseudotetragonal (Q) layers, e.g. lead sulfide-based (PbS) layers.[5] The complex Q layers contain four atomic layers of sulfide compounds with the formula MX, where M=$Sn^{2+}$, $Pb^{2+}$ or $Sb^{2+}$ and X=S. The H layer is a simple binary layer and consists of octahedrons of disulfide compounds with the formula $MX_2$, where M=$Sn^{4+}$ and $Fe^{2+}$ and X=S. We have presented schematically the crystal structure of the layered materials in Fig.1a. To probe the stoichiometry analysis, we employed energy-dispersive X-ray spectroscopy technique that results in an approximate chemical formula $Pb_6Sn_3Sb_3FeS_{18}$ (see the Supplementary Information), which falls within the generic chemical formula $(Pb,Sn)^{2+}_{6+x}Sb^{3+}_2Fe^{2+}Sn^{4+}_2S_{14+x}$, where $-1 \leq x \leq 0.25$ of materials from a mine in Bolivia.[6] This confirms the origin of our samples which are originated from a mine in San Jose, Bolivia.

Franckeite has interesting and complex electronic band structure.[7, 8] The band gap of the basic metal sulfides that are the basic ingredients of the franckeite layered structures ranges from 0.37 eV in PbS to 2.1 eV in $SnS_2$,[9] with franckeite itself having an effective infrared band-gap of 0.65 eV.[7, 8, 10] Hence franckeite provides an ideal platform to study band gap engineering,[11] optoelectronics, phase engineering[12] and thermoelectric properties[13] for the first time in a complex layered natural vdWH. Here we present our detail study of intrinsic opto-electronic behavior of the thin franckeite (60 nm < $d$ < 100 nm) for the first time. We have determined precisely the photocurrent (PC) responsivity, temperature dependence, power dependence and the time response of franckeite photodetectors. To understand our experimental measurements we have simulated the electronic band structure and the imaginary component of the dielectric constant via density function theory and obtained close agreement. Our study reveals fundamental opto-electronic properties of franckeite for the first time and can open up the possibilities of producing novel nanoscale optoelectronic devices with tailored properties.

The flakes of franckeite examined in this study were micro-exfoliated from a crust of franckeite crystals originating from San Jose, Bolivia. The scanning electron micrographs (SEM) of a sample on $SiO_2$/Si substrate is shown in Fig. 1b. The image demonstrates the layered nature of the flakes, which facilitates the micro-exfoliation process. We fabricated few layer electrically connected franckeite samples using the dry transfer technique and fabrication methods originally developed by Zomer et al.[14] (see Methods). The optical image of a franckeite device prepared on a glass substrate is shown in Fig. 1c. Thickness of the sample was measured by AFM (see Supplementary Information). We studied eleven samples and all devices demonstrated similar electrical and optical behavior.

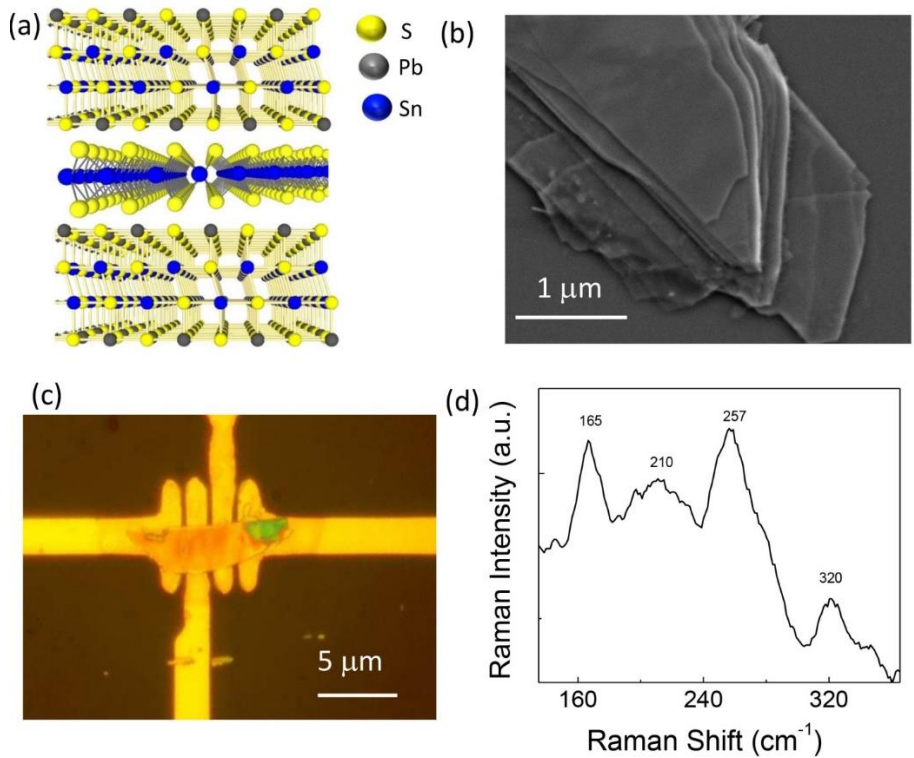

*Figure 1 : Franckeite crystal characterization. (a) Crystal structure of franckeite: the top and bottom layers are Q layers that includes MX compounds, where M=$Sn^{2+}$ or $Pb^{2+}$ (M can also be $Sb^{3+}$ instead of $Sn^{2+}$) and X=S. The middle layer or H-layer includes $MX_2$, where M=$Sn^{4+}$ (M can also be $Fe^{2+}$ instead of $Sn^{4+}$) and X=S. (b) Scanning electron micrographs (SEM) of a micro-exfoliated franckeite flake immobilized on a $SiO_2$/Si substrate. (c) The optical image of an electrically connected franckeite device. (d) Raman spectrum of a Franckeite. The laser excitation wavelength was 532 nm.*

The sample was further characterized by micro-Raman spectroscopy by pumping the sample with a 532 nm laser. We have observed peaks near 165, 210, 260 and 320 cm$^{-1}$ that are consistent with the recent report by Velicky *et al*[8]. The two peaks around 260 cm$^{-1}$ and 320 cm$^{-1}$ are probably originating from the stibnite ($Sb_2S_3$) and berndtite ($SnS_2$) in the H layer, respectively.[15-17] We also note that the 210 cm$^{-1}$ peak may also originate from the PbS lattice in the *H* layer.[18] The origin of the 165 cm$^{-1}$ is not currently known.

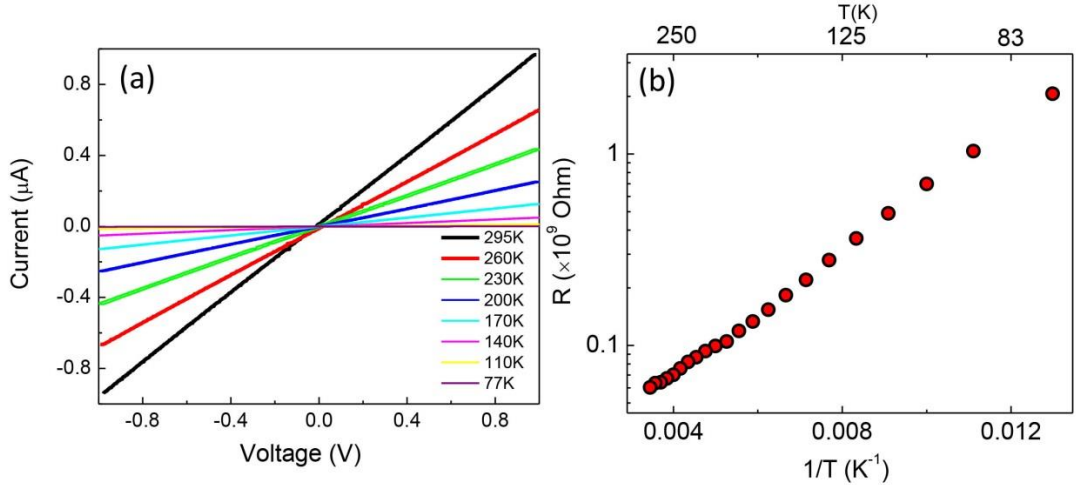

*Figure 2: Electrical measurement of franckeite crystals. (a) The I-V curve of a franckeite sample at different temperatures. (b) Logarithm of the resistance R versus inverse temperature 1/T for the I-V curve shown in (a). The linear fit yields $R \propto e^{E_A/2kT}$ with $E_A \sim$ 100 meV for the resistance measured at different temperatures.*

The electrical measurement at different temperatures was conducted with the aid of a microscopy cryostat. The measured two terminal current-voltage (*I-V*) curves at different temperatures are shown in Fig.2a. The *I-V* characteristics of the sample shown above shows excellent linear behavior suggesting the formation of Ohmic behavior. For some samples, we have also observed non-linear behavior at the origin, suggesting the formation of Schottky barrier at the Au and franckeite interface consistent with the findings reported by Velicky et al[8]. The logarithm of resistance for different temperatures is shown in Fig. 2b. We observed a clear Arrhenius behavior, $R \propto e^{E_A/2kt}$, with an activation energy $E_A$ ~100 meV. We argue that this activation energy is due to the dopant impurities in the sample,[8, 19] which introduce energy levels in the band gap. This argument is also supported by two recent reports of the observation of the *p*-doping of the franckeite.[7, 8]

We probed the photoresponse behavior of our franckeite samples by measuring temperature dependent photocurrent spectrum over a wide range of temperatures from 77K to room temperature (~295 K). The sample was mounted inside a microscopy cryostat (Janis Research, ST-500) equipped with electrical feedthrough for electro-optical measurements. We used a broadband light source (tungsten-halogen lamp) coupled to a double-grating monochromator (Acton spectra pro SP-2150i). The photocurrent was measured by employing lock-in techniques.[20] The photocurrent responsivity, photocurrent per unit power, of a representative sample is shown in Fig.3a.

Since we conducted broadband PC measurements from 300 nm to 2000 nm, we employed a double-grating monochromator to resolve the photon wavelength; 1) 600 g/mm with blazing-300 nm for the UV-VIS and 2) 600 g/mm with blazing-1000 nm for the infrared. For the PC measurement in the infrared regime, we used additional filter (cut-off 1100 nm) to remove the 2nd order visible light spectrum from the diffraction grating. The power of the optical spectrum was calibrated using a calibrated Si photodetector (Si PIN photodiode, Hamamatsu S1223, for the wavelength from 300 nm to 1100 nm) and an infrared photoconductive detector (PbS Photoconductor from Thorlabs for the wavelength from 1100nm to 2000 nm).

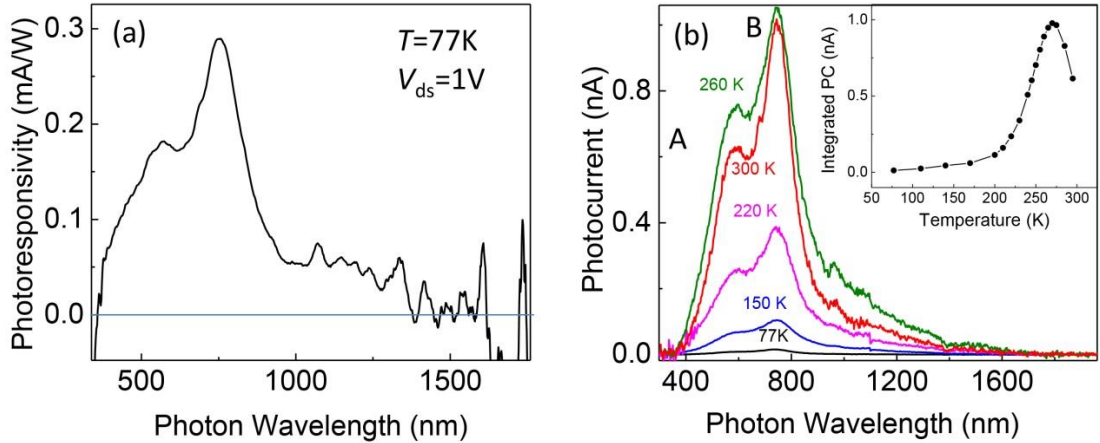

*Figure 3: Optoelectronic properties of franckeite thin film. (a) The PC spectrum of a franckeite device at 77K. The PC was normalized using a calibrated Si photodetector. The Photocurrent is diminishing at ~1450 nm indicating an effective band-gap of ~830 meV. (b) The PC spectrum for different temperature. The PC magnitude increases as we increase the temperature until ~260 K and then decrease again. Inset presents the integrated PC data as a function of the temperature showing the transition temperature ~260 K. The photocurrent integration was conducted between 400 nm and 800 nm. The overall shape PC spectrum remains the same.*

We measured the photoresponsivity of the sample, which is defined as $R=I_{ph}/P_{light}$, where $I_{ph}$ is the measured photocurrent and $P_{light}$ is the power of the incident light. Photoresponsivity is an important figure of merits for a photodetector.[20] The photoresponsivity data recorded at 77 K is shown in Fig. 3a. We have observed the maximum photoresponsivity 0.3 mA/W at ~750 nm at 77K. At room temperature the photoresponsivity is increased to 20 mA/W corresponding to an external quantum efficiency $EQE$~3%, which is very promising for franckeite based photonic device applications.

The PC spectrum for different temperature is presented in Fig.3b. We note three important features; (i) there are two peaks at 600 nm and 750 nm; (ii) the PC signal diminishes around 1500 nm (iii) the PC spectrum is very broad. The cut off at 1500 nm is originating from the band-edge, which confirms the narrow band gap of franckeite ($E_g$~ 830 meV). The energy of the two peaks, ~590 nm (~2.1 eV) and ~740 nm (~1.66 eV), are matching with the optical band gaps of $SnS_2$ and $Sb_2S_3$, respectively.[9] It is interesting to note that the photoresponse behavior of franckeite is dominated by $Sb_2S_3$. We also note that we observed these two peaks features only for three samples. For all other samples, we have observed only one peak at 1.9 eV (see Supplementary Information).

To understand the origin of these peaks in PC, we performed DFT calculations to determine the electronic structures of the franckeite material and to calculate the optical absorption properties. Fig.4a shows the calculated band structure of unit cell of bulk franckeite that contains 4 Pb, 6 Sn and 12 S atoms. The bulk franckeite shows finite density of states at the Fermi energy and there is a finite energy gap within the conduction band, which agrees with a recent calculation.[7] The frequency dependent dielectric function is calculated by a summation over empty states and plotted in Figure 4b and 4c. In addition, an isolated H, Q, HQ and HQH have been calculated for comparison. There is one major peak (Fig.

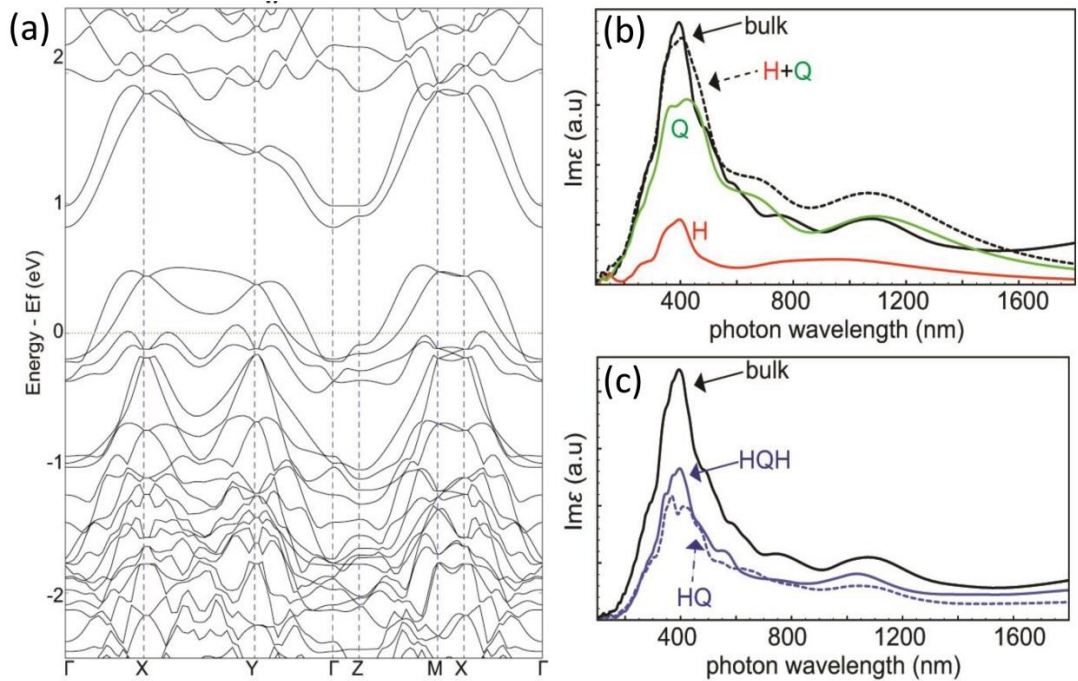

*Figure 4. Calculation of the optical absorption properties of franckeite. (a) DFT calculated band structure of the franckeite bulk material (b) The imaginary part of the dielectric function of an isolated H or Q layer as compared with the bulk franckeite. The dashed black line shows the sum over H and Q values. (c) The imaginary part of the dielectric function for the HQ bilayer (blue dashed) and HQH trilayer (blue solid) as compared with the bulk franckeite.*

4b) at short wavelengths, which is very similar to the sum over the individual H and Q layers (dashed black in Fig.4b). However, at the longer wavelength (above 800 nm), the difference between the bulk franckeite and sum over H and Q becomes more significant, indicating optical transition between these two materials at the interface may start to dominate. The long tail at the infrared wavelength in the calculated spectrum results from the excitation of electrons at the Fermi level to the empty states in the proximity of Fermi level. It should be noted that the calculated main peak is blue shifted as compared to the photocurrent measurement (Fig.3). This difference may result from the complicated distribution of the Pb and Sn atoms in the Q layer. However, Figure S3 in the supplement shows the calculated imaginary part of the dielectric function with Pb and Sn distributed in the Q layer with varied stacking configuration and these spectrums look very similar. Another possibility is the impurities in the franckeite materials as shown in the EDX measurement and in the literature.[5] In addition, it remains to be studied how the coulombic interaction between the excited electrons and holes will affect the optical spectrum calculations, since this interaction has not been included in the current calculations.

Now we explore the temperature dependence of PC and its origin. Temperature variation of the PC spectrum from a franckeite sample is shown in Fig.3b. We observed several interesting features. First, the PC magnitude is many orders lower at cryogenic temperature than room temperature. Second, the PC increases as we increase the temperature until it reaches 260K, then it starts to decreases again. To show this feature clearly, we have plotted integrated photocurrent (between 400 nm and 800 nm) as a function of the temperature as shown in the inset of Fig.3b. This temperature variation suggests that the PC is dominated by the trapping of charged carriers by mid-gap states.

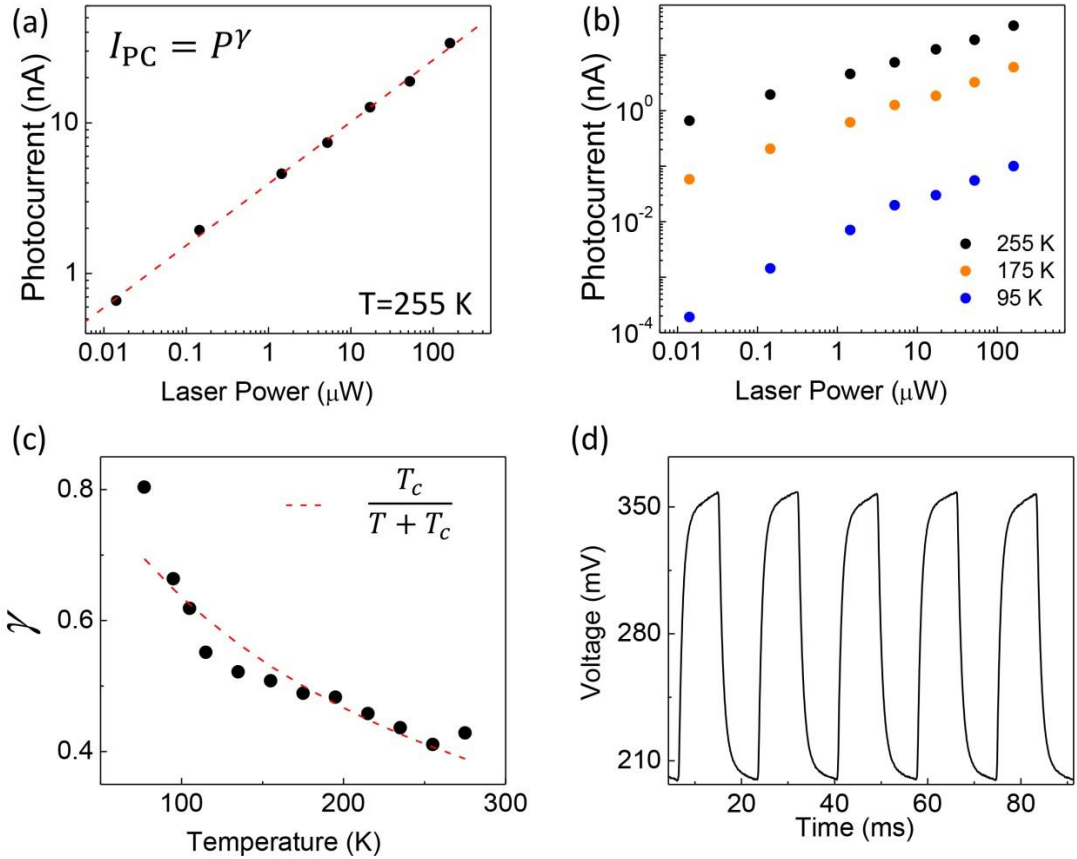

*Figure 5: Laser power dependence of the optoelectronic properties of franckeite and time response. (a) Variation of PC response of a franckeite device at 255K (applied voltage V=6V and laser wavelength 543 nm). PC follows a powe dependence, $I_{PC} = P^\gamma$. For the PC at 255K we found, $\gamma = 0.41$. (b) The PC data for different temperature in dependence of the low laser power. (c) Temperature dependence of the power exponent (γ) as a function of the temperature. The dashed line is a fit by an equation $T_c/(T+T_c)$, where $T_c \sim 175$ K. (d) The PC time response plot measured at room temperature. The sample was pumped by green laser and the response was measured by a high speed oscilloscope.*

Now we will explore the effect of the impurities and mid-gap states via power resolved PC measurement. The power dependence of the photocurrent at 255 K is shown in Fig.5a. The PC follows clearly a power dependence $I_{PC} = P^\gamma$ with $0.41 \leq \gamma \leq 0.8$ for different temperature as shown in Fig.5b-c. This non-linear power dependence can originate from several processes, such as photogating effect, thermoelectric effect and mid-gap trap states.[21] Since the device is residing on an optically-inactive glass substrate, we argue that there is no photogating effect[20] in our measured samples. Since the beam diameter of the sample is covering the whole sample uniformly, we argue that no temperature gradient exists to generate thermoelectric voltages. Hence, we can safely eliminate the thermoelectric effects determining the opto-electronic behavior.

Our temperature dependence of $\gamma$ suggests that the mid-gap states are responsible for the non-linear power dependence. We have conducted the power dependence at different temperature as shown in Fig.5b-c. The temperature variation of $\gamma$ at low laser power is related to the distribution of the localized trap states[21, 22] by a temperature dependence of the form $\gamma = \frac{T_c}{T+T_c}$, where $T_c$ is a characteristic temperature. The observed values of $\gamma$ are varying between 0.8 and 0.4 over a wide range of temperature (77 K <T < 275 K) and follow a similar temperature dependence with $T_c \sim 175$ K (dashed line in Fig.4c). This observed temperature dependence suggests that the trap states are continuously distributed below the conduction bands and above the valence bands.[21, 22] Hence our measurement confirms that the optoelectronic behavior of franckeite is dominated by continuous distribution of trap states.

Now we will focus on the time response behavior of the device. The time response of the photodetector is shown in Fig.5d. The laser was modulated by a pulse generator and the signal was measured by high speed oscilloscope. We have observed rising time ~0.4 ms and the drop-off time ~1.0 ms measured at room temperature. Hence the speed of our franckeite based photodetector is ~1KHz, suggesting that franckeite based device can be used in high speed applications.

In conclusion, we present the opto-electronic behavior of thin natural vdWH franckeite, which demonstrate a narrow band-gap semiconductor with a cut-off at ~1500 nm and broad photoresponse from UV to Infrared regime. By simulating the electronic band structure via DFT, we calculated the optical absorption properties, which is in agreement with the measured PC spectrum qualitatively. The devices demonstrate efficient photoresponse with external quantum efficiency *EQE*~3% and fast photoresponse behavior with response times as short as 1 ms (or speed ~1 KHz). Our temperature resolved and the power resolved PC measurements suggest that the photo-response behavior is dominated by trap states, which are acting as recombination centers and continuously distributed in the band gap. Our study shed light for the first time on the opto-electronic behavior of franckeite, a naturally occurring complex van der Waals heterostructure, advances fundamental understanding of its opto-electronic behaviors and can lead to the development new type of nanoscale optoelectronic devices with tailored properties.

**Methods:**

**Sample Fabrication:** <u>First</u>, we fabricated the metal electrodes on the glass substrate using optical lithography followed by thermal evaporation of Cr/Au[20] (5 nm/100 nm). In parallel, we prepared a PDMS (poly dimethyl-siloxane) stamp[23] structure on a microscope slide. The franckeite was exfoliated onto the PDMS stamp followed by characterization using optical microscopy, Raman spectroscopy and atomic force microscopy (AFM) to verify the franckeite deposition and determine the sample thickness (see Supplementary Information). We found that the sample thickness is varying from 60 nm to 800 nm. We aligned the exfoliated franckeite flake on the polymer stack with the pre-patterned metal electrode target under a microscope and brought it into contact. The polymer layer was mechanically separated from the PDMS stacks.

**Calculation Methods:** Density functional theory calculations were performed using the VASP package[24]. The PBE (Perdew-Burke-Ernzerhof) exchange-correlation potential[25] was used, and the electron-core interactions were treated in the projector augmented wave (PAW) method.[26, 27] The van der Waals interaction has been taken into account through the DFT-D3 semi-empirical methods.[28, 29] The calculations have been performed using a franckeite unit cell taken from an experimental measurement with the lattice constant a=5.805Å, b=5.856Å and c=17.338 Å.[5] The Brillouin zone of the bulk franckeite was sampled using a (6×6×2) k-point with a kinetic cut off energy of 400 eV. For the HQ and HQH layers, a (6×6×1) k-point was used for the calculations of optical absorption. The atomic structures were visualized using the VESTA software.[30]

**ACKNOWLEDGEMENT:**

AKMN, KR, SJ and AEY are grateful for the financial support from SFSU. KKHS and EP acknowledge support from the AFOSR grant FA9550-14-1-0251 and NSF EFRI 2-DARE grant 1542883. KKHS also acknowledges partial support from the Stanford Graduate Fellowship program and NSF Graduate Research Fellowship under Grant No. DGE-114747. TM and BW appreciate support from a NASA EPSCoR grant (NNX16AQ97A). This research used computational resources of the National Energy Research Scientific Computing Center, a DOE Office of Science User Facility supported by the Office of Science of the U.S. Department of Energy.

References:

1. Geim, A. K.; Grigorieva, I. V. Van der Waals heterostructures. *Nature* 2013, 499, 419-425.
2. Liu, Y.; Weiss, N. O.; Duan, X.; Cheng, H.-C.; Huang, Y.; Duan, X. Van der Waals heterostructures and devices. *Nature Reviews Materials* 2016, 1, 16042.
3. Novoselov, K. S.; Mishchenko, A.; Carvalho, A.; Castro Neto, A. H. 2D materials and van der Waals heterostructures. *Science* 2016, 353.
4. Butler, S. Z.; Hollen, S. M.; Cao, L.; Cui, Y.; Gupta, J. A.; Gutiérrez, H. R.; Heinz, T. F.; Hong, S. S.; Huang, J.; Ismach, A. F.; Johnston-Halperin, E.; Kuno, M.; Plashnitsa, V. V.; Robinson, R. D.; Ruoff, R. S.; Salahuddin, S.; Shan, J.; Shi, L.; Spencer, M. G.; Terrones, M.; Windl, W.; Goldberger, J. E. Progress, Challenges, and Opportunities in Two-Dimensional Materials Beyond Graphene. *Acs Nano* 2013, 7, 2898-2926.
5. Makovicky, E.; Petricek, V.; Dusek, M.; Topa, D. The crystal structure of franckeite, $Pb_{21.7}Sn_{9.3}Fe_{4.0}Sb_{8.1}S_{56.9}$. *Am. Mineral.* 2011, 96, 1686-1702.
6. Moh, G. H. Mutual Pb-2+/Sn-2+ Substitution in Sulfosalts. *Mineralogy and Petrology* 1987, 36, 191-204.
7. Molina-Mendoza, A. J.; Giovanelli, E.; Paz, W. S.; Niño, M. A.; Island, J. O.; Evangeli, C.; Aballe, L.; Foerster, M.; van der Zant, H. S. J.; Rubio-Bollinger, G.; Agraït, N.; Palacios, J. J.; Pérez, E. M.; Castellanos-Gomez, A. Franckeite as a naturally occurring van der Waals heterostructure. *Nature Communications* 2017, 8, 14409.
8. Velický, M.; Toth, P. S.; Rakowski, A. M.; Rooney, A. P.; Kozikov, A.; Woods, C. R.; Mishchenko, A.; Fumagalli, L.; Yin, J.; Zólyomi, V.; Georgiou, T.; Haigh, S. J.; Novoselov, K. S.; Dryfe, R. A. W. Exfoliation of natural van der Waals heterostructures to a single unit cell thickness. *Nature Communications* 2017, 8, 14410.
9. Xu, Y.; Schoonen, M. A. A. The absolute energy positions of conduction and valence bands of selected semiconducting minerals. *Am. Mineral.* 2000, 85, 543-556.
10. Boldish, S. I.; White, W. B. Optical band gaps of selected ternary sulfide minerals. *Am. Mineral.* 1998, 83, 865-871.
11. Unuchak, D. M.; Bente, K.; Kloess, G.; Schmitz, W.; Gremenok, V. F.; Ivanov, V. A.; Ukhov, V. Structure and optical properties of PbS-SnS mixed crystal thin films. *Physica Status Solidi C - Current Topics in Solid State Physics, Vol 6, No 5* 2009, 6, 1191-1194.
12. Soriano, R. B.; Malliakas, C. D.; Wu, J.; Kanatzidis, M. G. Cubic Form of $Pb_{2-x}Sn_xS_2$ Stabilized through Size Reduction to the Nanoscale. *J. Am. Chem. Soc.* 2012, 134, 3228-3233.

13. He, J.; Girard, S. N.; Zheng, J.-C.; Zhao, L.; Kanatzidis, M. G.; Dravid, V. P. Strong Phonon Scattering by Layer Structured PbSnS2 in PbTe Based Thermoelectric Materials. *Adv. Mater.* 2012, 24, 4440-4444.
14. Zomer, P. J.; Dash, S. P.; Tombros, N.; van Wees, B. J. A transfer technique for high mobility graphene devices on commercially available hexagonal boron nitride. *Appl. Phys. Lett.* 2011, 99, 232104.
15. Smith, A. J.; Meek, P. E.; Liang, W. Y. Raman scattering studies of $SnS_2$ and $SnSe_2$. *Journal of Physics C: Solid State Physics* 1977, 10, 1321.
16. Makreski, P.; Petruševski, G.; Ugarković, S.; Jovanovski, G. Laser-induced transformation of stibnite (Sb2S3) and other structurally related salts. *Vib. Spectrosc* 2013, 68, 177-182.
17. Mead, D. G.; Irwin, J. C. Raman spectra of SnS2 and SnSe2. *Solid State Commun.* 1976, 20, 885-887.
18. Nanda, K. K.; Sahu, S. N.; Soni, R. K.; Tripathy, S. Raman spectroscopy of PbS nanocrystalline semiconductors. *Physical Review B* 1998, 58, 15405-15407.
19. Tsang, W. T.; Schubert, E. F.; Cunningham, J. E. Doping in semiconductors with variable activation energy. *Appl. Phys. Lett.* 1992, 60, 115-117.
20. Klots, A. R.; Newaz, A. K. M.; Wang, B.; Prasai, D.; Krzyzanowska, H.; Lin, J.; Caudel, D.; Ghimire, N. J.; Yan, J.; Ivanov, B. L.; Velizhanin, K. A.; Burger, A.; Mandrus, D. G.; Tolk, N. H.; Pantelides, S. T.; Bolotin, K. I. Probing excitonic states in suspended two-dimensional semiconductors by photocurrent spectroscopy. *Sci. Rep.* 2014, 4, 6608.
21. Ghosh, S.; Winchester, A.; Muchharla, B.; Wasala, M.; Feng, S. M.; Elias, A. L.; Krishna, M. B. M.; Harada, T.; Chin, C.; Dani, K.; Kar, S.; Terrones, M.; Talapatra, S. Ultrafast Intrinsic Photoresponse and Direct Evidence of Sub-gap States in Liquid Phase Exfoliated MoS2 Thin Films. *Scientific Reports* 2015, 5.
22. Haynes, J. R.; Hornbeck, J. A. Temporary Traps in Silicon and Germanium. *Physical Review* 1953, 90, 152-153.
23. Xia Y., W. G. M. Soft Lithography. *Annu. Rev. Mater. Sci.* 1998, 28, 153-184.
24. Kresse, G.; Furthmuller, J. Efficient iterative schemes for ab initio total-energy calculations using a plane-wave basis set. *Phys Rev B* 1996, 54, 11169-11186.
25. Perdew, J. P.; Burke, K.; Ernzerhof, M. Generalized gradient approximation made simple. *Phys Rev Lett* 1996, 77, 3865-3868.
26. Blochl, P. E. Projector Augmented-Wave Method. *Phys Rev B* 1994, 50, 17953-17979.
27. Kresse, G.; Joubert, D. From ultrasoft pseudopotentials to the projector augmented-wave method. *Phys Rev B* 1999, 59, 1758-1775.
28. Grimme, S.; Antony, J.; Ehrlich, S.; Krieg, H. A consistent and accurate ab initio parametrization of density functional dispersion correction (DFT-D) for the 94 elements H-Pu. *J Chem Phys* 2010, 132, 154104.
29. Grimme, S.; Ehrlich, S.; Goerigk, L. Effect of the Damping Function in Dispersion Corrected Density Functional Theory. *J Comput Chem* 2011, 32, 1456-1465.
30. Momma, K.; Izumi, F. VESTA 3 for three-dimensional visualization of crystal, volumetric and morphology data. *J Appl Crystallogr* 2011, 44, 1272-1276.

# Supplementary Information

## Photoresponse of Natural van der Waals Heterostructures

Kyle Ray[¶], Alexander E. Yore[¶], Tong Mou[†], Sauraj Jha[¶], K.K.H. Smithe[‡], Bin Wang[†], Eric Pop[‡] and A.K.M. Newaz[¶]

[¶]Department of Physics and Astronomy, San Francisco State University, San Francisco, California 94132, United States,

[†]School of Chemical, Biological and Materials Engineering, University of Oklahoma, Norman, Oklahoma 73019.

[‡]Department of Electrical Engineering, Stanford University, Stanford, California 94305.

**Energy Dispersive Spectroscopy (EDXS) on the SEM of Franckeite**

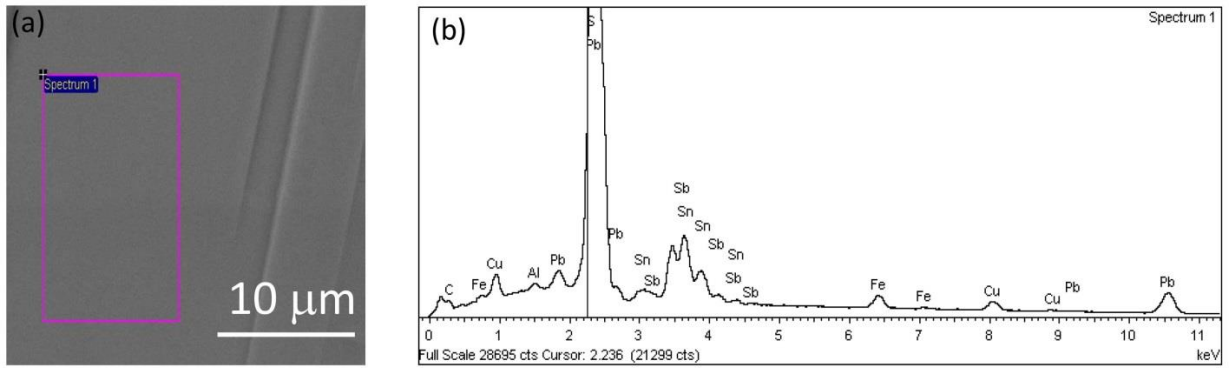

Figure S1: SEM and EDXS characterization of franckeite. (a) SEM image of a franckeite flake used for the EDXS study marked by pink square. (b) Average EDXS spectra obtained for the sample.

| Element | Atomic% |
|---|---|
| C K | 0.00 |
| Al K | 1.21 |
| S K | 54.89 |
| Fe K | 3.51 |
| Cu K | 3.84 |
| Sn L | 9.29 |
| Sb L | 9.14 |
| Pb M | 18.12 |
| Totals | 100 |

Table 1: EDXS quantification of franckeite crystal.

## Photoresponse of franckeite sample not showing the peak at 2.1 eV.

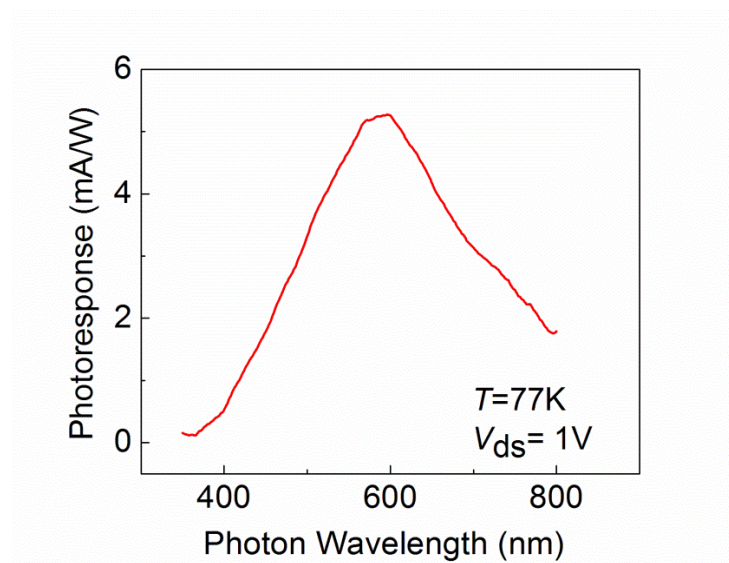

Figure S2: The PC response of franckeite crystal that demonstrates only one peak at ~1.9 eV. The measurement was conducted at 77K and the drain source voltage was $V_{ds}$=1V.

## DFT Calculations

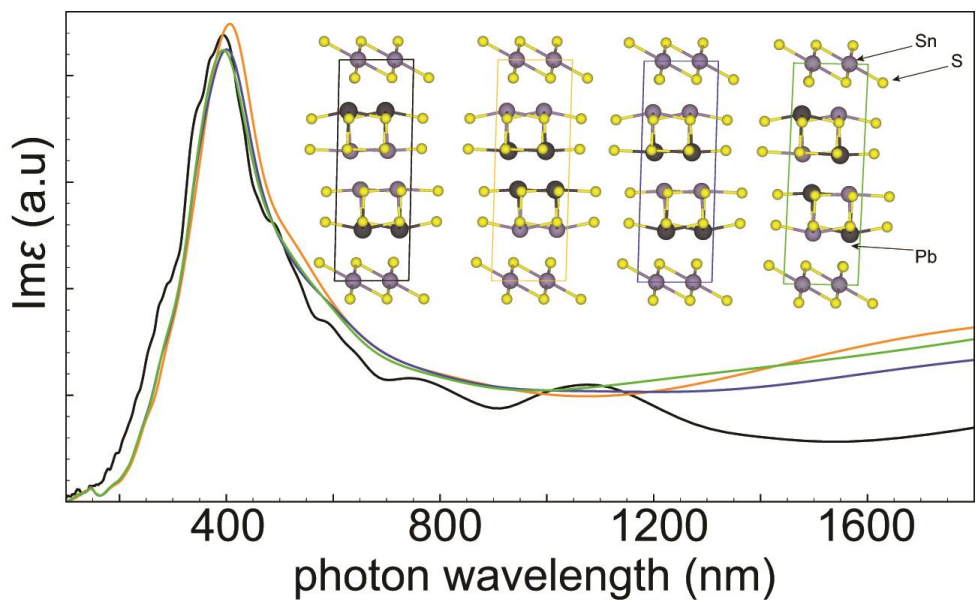

**Figure S3:** DFT-PBE calculated imaginary part of the dielectric function of the Franckeite bulk structure with varied stacking (insets) in the Q layer.

# Thickness measurement

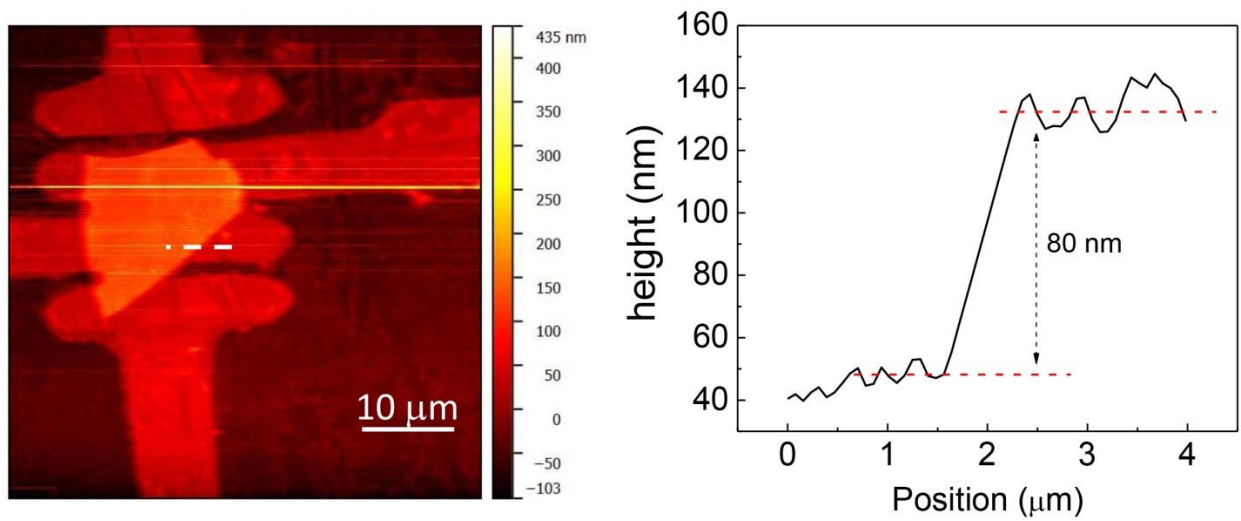

Figure S4: Thickness measurement by AFM. (Left) An AFM image of the flake on Au electrodes taken after the dry transfer. (b) The line profile of the sample along the white dashed line in the left figure.

4